\documentclass[prl,amssymb,twocolumn,showpacs,superscriptaddress,floatfix]{revtex4}

\usepackage{graphicx}
\usepackage{bm}
\usepackage{amsmath}

\usepackage{color}

\begin{document}

\title{Griffiths-like phase in the three-dimensional $\pm J$ Edwards-Anderson spin-glass model}

\author{M. L. Rubio Puzzo}
\affiliation{Instituto de Investigaciones Fisicoqu\'{i}micas Te\'oricas y Aplicadas (INIFTA), UNLP,CCT La Plata - CONICET, c.c. 16, Suc. 4, 1900 La Plata, Argentina}
\affiliation{CONICET, Centro At{\'{o}}mico Bariloche, 8400 San Carlos de
Bariloche, R\'{\i}o Negro, Argentina}
\author{F. Rom\'a}
\affiliation{CONICET, Departamento de F\'{\i}sica, Universidad Nacional de San Luis, D5700HHW San Luis, Argentina}
\author{S. Bustingorry}
\affiliation{CONICET, Centro At{\'{o}}mico Bariloche, 8400 San Carlos de
Bariloche, R\'{\i}o Negro, Argentina}
\author{P. M. Gleiser}
\affiliation{CONICET, Centro At{\'{o}}mico Bariloche, 8400 San Carlos de
Bariloche, R\'{\i}o Negro, Argentina}

\begin{abstract}
Using information of the ground state topology and the damage spreading technique we show a Griffiths-like phase is present in the three-dimensional $\pm J$ Edwards-Anderson spin glass model. With spin-flipping dynamics damage spreads for temperatures larger than $T_{\mathrm{g}}$, the glass transition temperature. With spin-orienting dynamics and for temperatures in $T_{\mathrm{g}} < T < T_{\mathrm{d}}$, damage spreads over a finite region of the system, composed of finite clusters of ferromagnetic character. $T_{\mathrm{d}}$ is the spin-orienting damage critical temperature, which is of the same order as the critical temperature of the ferromagnetic Ising model, $T_{\mathrm{c}}$. The present results allow us to identify the clusters which originate the Griffiths-like phase in the present model.
\end{abstract}


\date{\today}

\maketitle

The nature of the low-temperature phase of spin glasses has been an elusive issue over many years despite intensive studies~\cite{Young}. It is now well established that there exists a spin glass phase below a characteristic temperature $T_{\mathrm{g}}$ for Ising spin glasses, and that far above $T_{\mathrm{g}}$ the system is in a paramagnetic state. The critical temperature $T_{\mathrm{g}}$ has been numerically calculated using diverse methods, most of them adopting the idea of a \textit{global} growing characteristic length in the spin glass phase (see Ref.~\cite{katzgraber2006} and references therein). However, it is not clear what is the origin of the `order' behind the growing length. What is the underlying structure of the low-temperature spin glass phase also remains an open subject. A different approach attempted to identify \textit{local} structures presenting typical features of ordered states.
These ideas have been further extended in the last decade through a systematic study of spatial and dynamical heterogeneities in spin glasses~\cite{Chamon02-etal,Castillo02-etal,Montanari03a}. In particular, the origin of strong dynamical heterogeneities is unveiled by ground state (GS) spatial heterogeneities in the $\pm J$ Edwards-Anderson (EA) spin glass model\cite{Roma2,Roma3-etal}.

The slow relaxation properties observed above $T_{\mathrm{g}}$ have also attracted much attention. Randeria \textit{et al.}~\cite{randeria1985} discussed the high temperature relaxation of spin glasses. They found that, for models with a bounded bond distribution, for temperatures between $T_{\mathrm{g}}$ and $T_{\mathrm{c}}$, the critical temperature of the orderer system, the time correlation function has a lower bound with a relaxation slower than exponential but faster than a power-law. This behavior was intuitively associated with a ``Griffiths-like'' phase. Let us recall that the Griffits phase was originally proposed for the site diluted Ising ferromagnetic model, where a fraction $1-p$ of the sites are removed~\cite{griffiths1969}. In the phase diagram of this system the ferromagnetic phase is present for temperatures smaller that $T_{\mathrm{c}}(p)$ up to the critical percolation value $p_{\mathrm{c}}$. It was argued that for temperatures in the range $T_{\mathrm{c}}(p) < T < T_{\mathrm{c}(p=1)}$, singularities in the zero field susceptibility and anomalous relaxation are observed due to the presence of finite size ferromagnetic clusters. Randeria \textit{et al.}~\cite{randeria1985} did not identify unfrustrated cluster corresponding to the Griffiths phase. Also, further numerical evidence does not agree with the predicted functional form analytically obtained considering Griffiths singularities~\cite{ogielski1985}. It has also been argued that these singularities are too weak to be observed in classical systems~\cite{Harris1975}. Extensions to quantum models, where the singularities are stronger, have been intensively studied~\cite{Guo1996}.

Further information on the dynamical properties above $T_{\mathrm{g}}$ has been obtained using the damage-spreading technique~\cite{hinrichsen2000}. In this case, one measures the configurational distance between two replicas of a system, which are initially set to different initial conditions, and evolve under the same thermal noise. Damage-spreading with spin-orienting dynamics in the three-dimensional (3D) $\pm J$ EA spin glass model was initially studied by Derrida and Weisbuch~\cite{derrida1987}. They found three temperature regimes: (i) a high-temperature regime $T>T_1 \simeq 4.1$ where the damage is zero, (ii) an intermediate regime $T_2\simeq 1.8<T<T_1$ where the damage is nonzero and independent of the initial condition, and (iii) a low-temperature regime $T<T_2$ where the damage depends on the initial condition. These authors stated that the determination of $T_1$ and $T_2$ could be consistent with $T_1=T_{\mathrm{c}}$ and $T_2=T_{\mathrm{g}}$. Note that this was a surprising result since there is no reason \textit{a priori} why an imprint of ferromagnetic order in the dynamical properties of this spin glass model should be expected.

Throughout this work, by using the damage-spreading technique with both spin-flipping and spin-orienting dynamical rules, and its correlation with the GS topological properties, we studied the dynamical properties of the 3D $\pm J$ EA model. This enables us to show that there is an ordered structure which presents ferromagnetic-like clusters capable of explaining both the anomalous relaxation and the ferromagnetic signature on the intermediate temperature range, $T_{\mathrm{g}} < T < T_{\mathrm{c}}$. Thus, these results give a natural framework to justify the presence of a Griffiths-like phase in the 3D $\pm J$ EA model. We also stress here that our aim is not to list the dynamical properties of this model with the damage-spreading technique, which have been extensively analyzed in the past. Instead, we seek for the constrained structure of the GS which unveils the Griffiths-like phase in the present model, and thus permits us to understand all the previous puzzling results within a single framework.

We consider, as a starting point, the topological properties of the GS configurations of the $\pm J$ EA model~\cite{Toulouse,Vannimenus,Barahona-etal,Ramirez1997,Roma2006-etal}. A systematic study has shown that it is possible to find a backbone, characterized by  solidary spins, which maintain their relative orientation over all the set of multiple-degenerate GS configurations~\cite{Barahona-etal,Roma2006-etal,Roma2009a}. In 3D, the subset of solidary spins comprises a fraction of approximately $76\%$ of the total spins. The cluster size distribution presents the same characteristics than in the diluted problem, i.e. a power-law decay form~\cite{Roma2009a}. Also, the largest cluster of the backbone percolates through the system. Besides, the backbone has a very small frustration (of the order of $10\%$ of the bonds), which suggests that ferromagnetic-like order can be sustained at low temperatures~\cite{Roma2009a}. Therefore, these characteristics naturally point to the backbone as the responsible for the existence of the Griffiths-like phase in spin glasses in the range $T_{\mathrm{g}} < T < T_{\mathrm{c}}$. For temperatures below $T_{\mathrm{g}}$ a ferromagnetic-like order grows within the largest cluster of the backbone~\cite{Roma2009c}. Another important characteristic is the presence of non-solidary spins, defined as those outside the backbone. These spins remain in a paramagnetic state for all temperatures.

\begin{figure}[!tbp]
\includegraphics[width=8cm,clip=true]{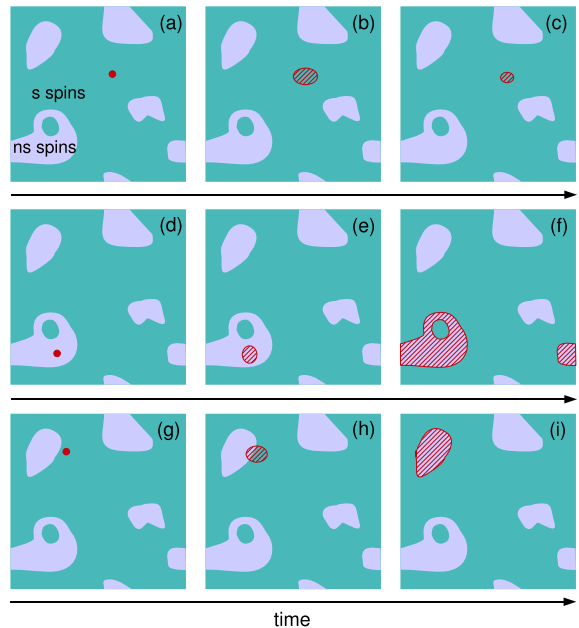}
\caption{\label{fig:picture} (Color online) Schematic representation of spin-flipping damage spreading (dashed-red) below $T_{\mathrm{c}}$ and considering the separation on solidary (green region) and non-solidary (violet region) spins. Each set of snapshots corresponds to the evolution of the system when the initial damage (red dot) is within: (a-c) the set of solidary spins. (d-f) an island of non-solidary spins. (g-i) the set of solidary spins but close to an island of non-solidary spins.}
\end{figure}

In the following we will show how damage-spreading technique can be used to unveil the limits of the Griffiths-like phase. It is well known that for the pure ferromagnetic model the results of this technique depend on the Monte Carlo dynamics~\cite{hinrichsen2000}. For spin-orienting dynamics the initial damage is healed above a damage temperature $T_{\mathrm{d}}=4.3$~\cite{lecaer1989,lecaer1989b}, and it spreads to a finite fraction of the sample below $T_{\mathrm{d}}$. The opposite scenario is observed when using a spin-flipping dynamics. The damage temperature $T_{\mathrm{d}}$ obtained with each dynamics is of the same order, and slightly lower than the critical temperature $T_{\mathrm{c}}=4.515$~\cite{binder2001}. Figure~\ref{fig:picture} shows a schematic view of a $\pm J$ EA system split up into solidary (green region) and non-solidary (violet region) spins. It is also shown how we expect it would respond to the location of an initial damage at a finite temperature $T<T_{\mathrm{g}}=1.12$~\cite{katzgraber2006} when a spin-flipping dynamics is considered. Figures~\ref{fig:picture}(a-c) qualitatively represent snapshots of the evolution of the system when the initial damage (red dot) is in the set of solidary spins, with the dashed red regions indicating damage spreading. Since the damaged region sustains a ferromagnetic-like order, the damage should rapidly heals. Instead, if the initial damage is located within an island of the non-percolating set of non-solidary spins, as represented in Figs.~\ref{fig:picture}(d-f), we expect damage to spread over the entire island reaching a finite fraction of the system. An interesting particular case is when the initial damage is within the set of solidary spins, but close to an island of non-solidary spins. In this case a fluctuation can allow the damage to reach the border and spread over all the island, as represented in Fig.~\ref{fig:picture}(g-i). When $T>T_{\mathrm{g}}$, since $T$ is above the ordering temperature of the largest component of the backbone, a large fraction of the system will be in a paramagnetic state and the small initial damage will propagate to the whole system.

The situation is different for spin-orienting dynamics. Above $T_{\mathrm{c}}$ all the system is paramagnetic and the damage does not propagate. For $T_{\mathrm{g}} < T < T_{\mathrm{c}}$ the largest cluster of the backbone is above its ordering temperature and thus the contribution of ferromagnetic-like order comes \textit{only} from the finite clusters of solidary spins, that are characterized by a power-law size distribution. Thus, these are the clusters over which damage propagates. Therefore, the damage-spreading technique over the finite temperature range $T_{\mathrm{g}} < T < T_{\mathrm{c}}$ is expected to uncover the presence of the unfrustrated clusters which have the structure to support a Griffiths-like phase. In the following, through exhaustive numerical simulations of the damage spreading process, we shall show quantitative results that support this picture.

The  $3D \pm J$ EA spin glass model is described through the Hamiltonian $H=\sum_{\langle i,j \rangle} J_{ij} \sigma_i \sigma_j$, where the sum runs over the nearest neighbors of a cubic lattice with linear size $L$ and $N=L^3$ spins, $\sigma_i=\pm 1$ are spin variables, and the random bonds $J_{ij}=\pm J$ are chosen from a symmetric bimodal distribution. The damage spreading protocol is as follows: for a given disorder realization we build two replicas of the system, $\alpha$ and $\beta$, one resulting from equilibration at a given temperature and the other from the same configuration but with a single damaged spin, i.e. with a randomly selected spin that is reversed. Equilibrium configurations were obtained with a parallel tempering Monte Carlo algorithm\cite{Hukushima1996}. The two replicas of the system evolve under the same thermal noise, i.e. using the same sequence of random numbers, and the Hamming distance between each sample is measured. Thus, the total damage at a given time $t$ is defined as
\begin{equation}
 \label{eq:damage-tot}
 D(t) = \frac{1}{4N} \sum_i \overline{\left\langle \left( \sigma_i^\alpha - \sigma_i^\beta\right)^2 \right\rangle},
\end{equation}
where the angular brackets and the overline correspond to averages over thermal histories and disorder realizations, respectively. In this way the damage can take values in the range $0 \leq D(t) \leq 1$. In the following we will present data for cubic geometry with $L=8$, where reliable information of GS configurations can be obtained~\cite{Roma2009a}. The disorder average is taken over $1000$ samples, for which we have determined the backbone structure through extensive numerical simulations. Whenever possible we will compare with a larger system size in order to show that our data does not suffer from strong finite size effects.

\begin{figure}[!tbp]
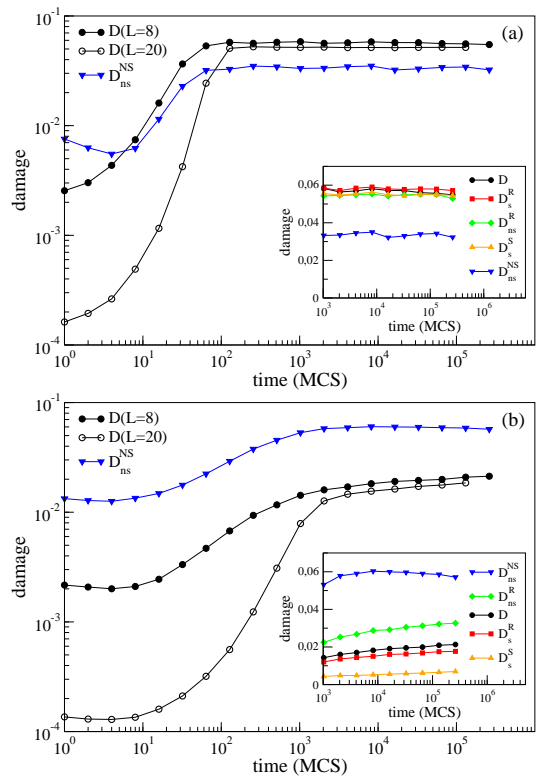

\includegraphics[width=7cm,clip=true]{fig2a.eps}
\includegraphics[width=7cm,clip=true]{fig2b.eps}
\caption{\label{fig:HB}  (Color online) Evolution of the damage with (a) spin-orienting dynamics at $T=2.5$ and (b) spin-flipping dynamics at $T=0.6$, comparing the total damage $D$ with $D_{\mathrm{ns}}^{\mathrm{S}}$. The total damage for $L=20$ is also shown. The inset shows data in linear scale for the stationary regime.}
\end{figure}

We begin our analysis considering the spin-orienting heat-bath rule. At high temperatures the total damage for the spin glass model vanishes at long times, while at low temperatures it arrives at a stationary regime. Figure~\ref{fig:HB}(a) shows the time evolution of the total damage at $T=2.5 < T_{\mathrm{d}}$ for $L=8$ (closed circles) and $L=20$ (open circles). The initial damage is $D(0)=1/N$, which corresponds to a flip of a random site. Note that for both system sizes the damage tends to almost the same stationary value, showing that there are no important finite size effects. 

Now, for a given time evolution of the whole system, we also perform a constrained measure of the damage restricting the sum in Eq. \eqref{eq:damage-tot} to solidary spins, denoted as $D_{\mathrm{s}}$, or to non-solidary spins, denoted as $D_{\mathrm{ns}}$. Notice that the normalization factor and the initial damage will be thus given by the average number of solidary or non-solidary spins, $1/N_{\mathrm{s}}$ and $1/N_{\mathrm{ns}}$, respectively. Furthermore, as one can exactly locate the initial damage, it is possible to have different constrained measures such as: (i) the damage in solidary spins when the initial damage is in any random spin, which we shall denote as $D_{\mathrm{s}}^{\mathrm{R}}$, (ii) the damage in non-solidary spins when the initial damage is in any random spin, $D_{\mathrm{ns}}^{\mathrm{R}}$, (iii) the damage in solidary spins when the initial damage is within the set of solidary spins, $D_{\mathrm{s}}^{\mathrm{S}}$, (iv) the damage in non-solidary spins when the initial damage is within an island of non-solidary spins, $D_{\mathrm{ns}}^{\mathrm{NS}}$, etc. In Fig.~\ref{fig:HB}(a) we show the constrained measure $D_{\mathrm{ns}}^{\mathrm{NS}}$. The inset shows the other constrained measures as indicated. It can be clearly observed that the only curve which is appreciably different in the stationary regime is $D_{\mathrm{ns}}^{\mathrm{NS}}$, which is smaller than the other curves. Instead, the damage can spread over the finite clusters of solidary spins below $T_{\mathrm{d}}$, resulting in $D > D_{\mathrm{ns}}^{\mathrm{NS}}$.

We also consider the spin-flipping Metropolis rule. Figure~\ref{fig:HB}(b) shows the total damage with this dynamics at $T=0.6<T_{\mathrm{g}}$,  for $L=8$ and $L=20$, and also its comparison with $D_{\mathrm{ns}}^{\mathrm{NS}}$. After a transient regime all the curves tend to stationary values. The inset shows the other constrained measures. With spin-flipping dynamics damage cannot spread over the backbone of solidary spins since they are far below their critical temperature, as depicted in Fig.~\ref{fig:picture}. However, damage spreading can be observed for non-solidary spins. One can thus observe that $D_{\mathrm{ns}}^{\mathrm{R}} > D_{\mathrm{s}}^{\mathrm{R}}$, i.e. no matter where the initial damage is, damage spreads mostly over non-solidary spins. The difference is even larger between $D_{\mathrm{ns}}^{\mathrm{NS}}$ and $D_{\mathrm{s}}^{\mathrm{S}}$.

\begin{figure}[!tbp]
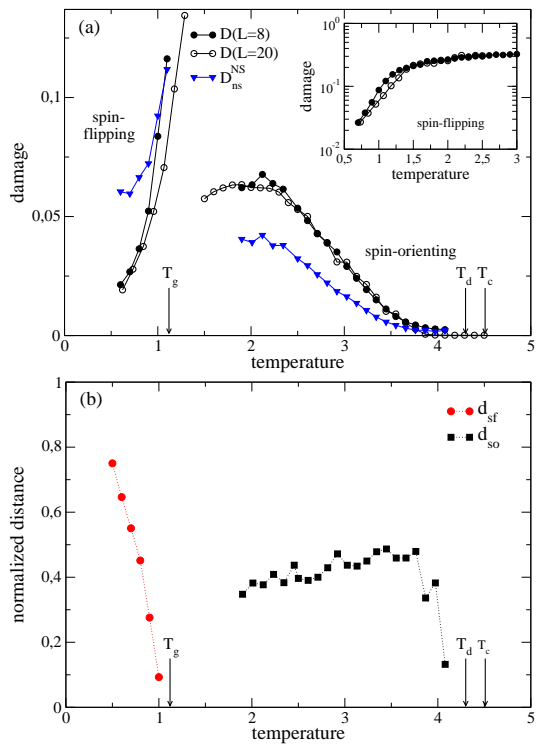

\includegraphics[width=7cm,clip=true]{fig3a.eps}
\includegraphics[width=7cm,clip=true]{fig3b.eps}
\caption{\label{fig:tgytc} (Color online)(a) Temperature dependence of the damage for spin-flipping and spin-orienting dynamics. The total damage for $L=20$ is also shown. The inset shows a comparison between $L=8$ and $L=20$ for spin-flipping dynamics in an enlarged scale. (b) Normalized distances between $D$ and $D_{\mathrm{ns}}^{\mathrm{NS}}$ (see the text).}
\end{figure}

The temperature dependence of the stationary damage is presented in Fig.~\ref{fig:tgytc}(a). We present the behavior of the total damage for $L=8$ and $L=20$ and also $D_{\mathrm{ns}}^{\mathrm{NS}}$ for $L=8$. Note that the temperature scales used for spin-flipping or spin-orienting dynamics are different. Both for spin-orienting and spin-flipping dynamics, the measure of the damage spreading restricted to non-solidary spins when the initial damage is in the subset of non-solidary spins, $D_{\mathrm{ns}}^{\mathrm{NS}}$, gives a distinctive feature of the non-trivial separation originated in the GS topology of the system. This can be quantified by a normalized distance between $D_{\mathrm{ns}}^{\mathrm{NS}}$ and $D$ defined as
\begin{equation}
 d_{\mathrm{sf}} = \frac{D_{\mathrm{ns}}^{\mathrm{NS}} - D}{D_{\mathrm{ns}}^{\mathrm{NS}}}, \qquad
 d_{\mathrm{so}} = \frac{D-D_{\mathrm{ns}}^{\mathrm{NS}}}{D}
\end{equation}
where  $d_{\mathrm{sf}}$  and $d_{\mathrm{so}}$ correspond to spin-flipping and spin-orienting dynamics, respectively. The temperature dependence of $d_{\mathrm{so}}$ and $d_{\mathrm{sf}}$ is presented in Fig.~\ref{fig:tgytc}(b), together with the relevant temperature scales, $T_{\mathrm{g}}$, $T_{\mathrm{d}}$ and $T_{\mathrm{c}}$.

We show here that spin-orienting and spin-flipping damage-spreading processes in spin glasses do not have the same damage spreading temperature as observed in ferromagnets. Indeed, in Fig.~\ref{fig:tgytc}(b) one can observe that $d_{\mathrm{sf}}$ is larger than zero below a temperature of the order of $T_{\mathrm{g}}$. This result indicates the ferromagnetic-like character of the largest cluster of the backbone below $T_{\mathrm{g}}$. On the other hand, $d_{\mathrm{so}}$ signals an important difference between solidary and non-solidary spins close to the Ising damage spreading temperature $T_{\mathrm{d}}$. This difference is directly related to the ferromagnetic character of the finite sized clusters of solidary spins. Indeed, this gives a reliable picture allowing to understand the puzzling question on the origin of both Griffiths-like anomalies and ferromagnetic signatures for $T_{\mathrm{g}} < T < T_{\mathrm{c}}$~\cite{Harris1975,randeria1985,campbell1994,jan1994,Guo1996}. In fact, the GS information, used to identify the clusters which originate the Griffiths-like phase in the present classical problem, gives an important insight to rationalize the presence of \textit{rare clusters} related to the origin of the Griffiths phase in the quantum transverse-field Ising spin glass.

\acknowledgments
FR acknowledges financial support from  projects PICT05-33328 and PICT07-2185, and U.N. de San Luis under project 322000. MLRP thanks INTER-U program (SPU-ME) for financial support.

\bibliography{spinglass}

\begin{thebibliography}{27}
\expandafter\ifx\csname natexlab\endcsname\relax\def\natexlab#1{#1}\fi
\expandafter\ifx\csname bibnamefont\endcsname\relax
  \def\bibnamefont#1{#1}\fi
\expandafter\ifx\csname bibfnamefont\endcsname\relax
  \def\bibfnamefont#1{#1}\fi
\expandafter\ifx\csname citenamefont\endcsname\relax
  \def\citenamefont#1{#1}\fi
\expandafter\ifx\csname url\endcsname\relax
  \def\url#1{\texttt{#1}}\fi
\expandafter\ifx\csname urlprefix\endcsname\relax\def\urlprefix{URL }\fi
\providecommand{\bibinfo}[2]{#2}
\providecommand{\eprint}[2][]{\url{#2}}

\bibitem[{\citenamefont{Young}(1997)}]{Young}
\bibinfo{author}{\bibfnamefont{A.~P.} \bibnamefont{Young}},
  \emph{\bibinfo{title}{Spin Glasses and Random Fields}}
  (\bibinfo{address}{Singapore}, \bibinfo{year}{1997}), \bibinfo{edition}{world
  scientific} ed.

\bibitem[{\citenamefont{Katzgraber et~al.}(2006)\citenamefont{Katzgraber,
  K\"orner, and Young}}]{katzgraber2006}
\bibinfo{author}{\bibfnamefont{H.~G.} \bibnamefont{Katzgraber}},
  \bibinfo{author}{\bibfnamefont{M.}~\bibnamefont{K\"orner}}, \bibnamefont{and}
  \bibinfo{author}{\bibfnamefont{A.~P.} \bibnamefont{Young}},
  \bibinfo{journal}{Phys. Rev. B} \textbf{\bibinfo{volume}{73}},
  \bibinfo{pages}{224432} (\bibinfo{year}{2006}).

\bibitem[{\citenamefont{Chamon et~al.}(2002)}]{Chamon02-etal}
\bibinfo{author}{\bibfnamefont{C.}~\bibnamefont{Chamon}} \bibnamefont{et~al.},
  \bibinfo{journal}{Phys. Rev. Lett.} \textbf{\bibinfo{volume}{89}},
  \bibinfo{pages}{217201} (\bibinfo{year}{2002}).

\bibitem[{\citenamefont{Castillo et~al.}(2002)}]{Castillo02-etal}
\bibinfo{author}{\bibfnamefont{H.~E.} \bibnamefont{Castillo}}
  \bibnamefont{et~al.}, \bibinfo{journal}{Phys. Rev. Lett.}
  \textbf{\bibinfo{volume}{88}}, \bibinfo{pages}{237201}
  (\bibinfo{year}{2002}).

\bibitem[{\citenamefont{Montanari and Ricci-Tersenghi}(2003)}]{Montanari03a}
\bibinfo{author}{\bibfnamefont{A.}~\bibnamefont{Montanari}} \bibnamefont{and}
  \bibinfo{author}{\bibfnamefont{F.}~\bibnamefont{Ricci-Tersenghi}},
  \bibinfo{journal}{Phys. Rev. Lett.} \textbf{\bibinfo{volume}{90}},
  \bibinfo{pages}{017203} (\bibinfo{year}{2003}).

\bibitem[{\citenamefont{Rom\'a et~al.}(2006{\natexlab{a}})\citenamefont{Rom\'a,
  Bustingorry, and Gleiser}}]{Roma2}
\bibinfo{author}{\bibfnamefont{F.}~\bibnamefont{Rom\'a}},
  \bibinfo{author}{\bibfnamefont{S.}~\bibnamefont{Bustingorry}},
  \bibnamefont{and} \bibinfo{author}{\bibfnamefont{P.~M.}
  \bibnamefont{Gleiser}}, \bibinfo{journal}{Phys. Rev. Lett.}
  \textbf{\bibinfo{volume}{96}}, \bibinfo{pages}{167205}
  (\bibinfo{year}{2006}{\natexlab{a}}).

\bibitem[{\citenamefont{Rom\'a et~al.}(2007)}]{Roma3-etal}
\bibinfo{author}{\bibfnamefont{F.}~\bibnamefont{Rom\'a}} \bibnamefont{et~al.},
  \bibinfo{journal}{Phys. Rev. Lett.} \textbf{\bibinfo{volume}{98}},
  \bibinfo{pages}{097203} (\bibinfo{year}{2007}).

\bibitem[{\citenamefont{Randeria et~al.}(1985)\citenamefont{Randeria, Sethna,
  and Palmer}}]{randeria1985}
\bibinfo{author}{\bibfnamefont{M.}~\bibnamefont{Randeria}},
  \bibinfo{author}{\bibfnamefont{J.~P.} \bibnamefont{Sethna}},
  \bibnamefont{and} \bibinfo{author}{\bibfnamefont{R.~G.}
  \bibnamefont{Palmer}}, \bibinfo{journal}{Phys. Rev. Lett.}
  \textbf{\bibinfo{volume}{54}}, \bibinfo{pages}{1321} (\bibinfo{year}{1985}).

\bibitem[{\citenamefont{Griffiths}(1969)}]{griffiths1969}
\bibinfo{author}{\bibfnamefont{R.~B.} \bibnamefont{Griffiths}},
  \bibinfo{journal}{Phys. Rev. Lett.} \textbf{\bibinfo{volume}{23}},
  \bibinfo{pages}{17} (\bibinfo{year}{1969}).

\bibitem[{\citenamefont{Ogielski}(1985)}]{ogielski1985}
\bibinfo{author}{\bibfnamefont{A. T.}~\bibnamefont{Ogielski}},
  \bibinfo{journal}{Phys. Rev. B} \textbf{\bibinfo{volume}{32}},
  \bibinfo{pages}{7384} (\bibinfo{year}{1985}).

\bibitem[{\citenamefont{Harris}(1975)}]{Harris1975}
\bibinfo{author}{\bibfnamefont{A.~B.} \bibnamefont{Harris}},
  \bibinfo{journal}{Phys. Rev. B} \textbf{\bibinfo{volume}{12}},
  \bibinfo{pages}{203} (\bibinfo{year}{1975}).

\bibitem[{\citenamefont{Guo et~al.}(1996)\citenamefont{Guo, Bhatt, and
  Huse}}]{Guo1996}
\bibinfo{author}{\bibfnamefont{M.}~\bibnamefont{Guo}},
  \bibinfo{author}{\bibfnamefont{R.~N.} \bibnamefont{Bhatt}}, \bibnamefont{and}
  \bibinfo{author}{\bibfnamefont{D.~A.} \bibnamefont{Huse}},
  \bibinfo{journal}{Phys. Rev. B} \textbf{\bibinfo{volume}{54}},
  \bibinfo{pages}{3336} (\bibinfo{year}{1996}).

\bibitem[{\citenamefont{Hinrichsen}(2000)}]{hinrichsen2000}
\bibinfo{author}{\bibfnamefont{H.}~\bibnamefont{Hinrichsen}},
  \bibinfo{journal}{Adv. Phys.} \textbf{\bibinfo{volume}{49}},
  \bibinfo{pages}{815} (\bibinfo{year}{2000}).

\bibitem[{\citenamefont{Derrida and Weisbuch}(1987)}]{derrida1987}
\bibinfo{author}{\bibfnamefont{B.}~\bibnamefont{Derrida}} \bibnamefont{and}
  \bibinfo{author}{\bibfnamefont{G.}~\bibnamefont{Weisbuch}},
  \bibinfo{journal}{Europhys. Lett.} \textbf{\bibinfo{volume}{4}},
  \bibinfo{pages}{657} (\bibinfo{year}{1987}).

\bibitem[{\citenamefont{Toulouse}(1977)}]{Toulouse}
\bibinfo{author}{\bibfnamefont{G.}~\bibnamefont{Toulouse}},
  \bibinfo{journal}{Commun. Phys.} \textbf{\bibinfo{volume}{2}},
  \bibinfo{pages}{115} (\bibinfo{year}{1977}).

\bibitem[{\citenamefont{Vannimenus et~al.}(1979)\citenamefont{Vannimenus,
  Maillard, and de~S\`eze}}]{Vannimenus}
\bibinfo{author}{\bibfnamefont{J.}~\bibnamefont{Vannimenus}},
  \bibinfo{author}{\bibfnamefont{J.~M.} \bibnamefont{Maillard}},
  \bibnamefont{and}
  \bibinfo{author}{\bibfnamefont{L.}~\bibnamefont{de~S\`eze}},
  \bibinfo{journal}{J. Phys. C: Solid State Phys.}
  \textbf{\bibinfo{volume}{12}}, \bibinfo{pages}{4523} (\bibinfo{year}{1979}).

\bibitem[{\citenamefont{Barahona et~al.}(1982)}]{Barahona-etal}
\bibinfo{author}{\bibfnamefont{F.}~\bibnamefont{Barahona}}
  \bibnamefont{et~al.}, \bibinfo{journal}{J. Phys. A}
  \textbf{\bibinfo{volume}{15}}, \bibinfo{pages}{673} (\bibinfo{year}{1982}).

\bibitem[{\citenamefont{Ramirez-Pastor
  et~al.}(1997)\citenamefont{Ramirez-Pastor, Nieto, and Vogel}}]{Ramirez1997}
\bibinfo{author}{\bibfnamefont{A.~J.} \bibnamefont{Ramirez-Pastor}},
  \bibinfo{author}{\bibfnamefont{F.}~\bibnamefont{Nieto}}, \bibnamefont{and}
  \bibinfo{author}{\bibfnamefont{E.~E.} \bibnamefont{Vogel}},
  \bibinfo{journal}{Phys. Rev. B} \textbf{\bibinfo{volume}{55}},
  \bibinfo{pages}{14323} (\bibinfo{year}{1997}).

\bibitem[{\citenamefont{Rom\'a et~al.}(2006{\natexlab{b}})}]{Roma2006-etal}
\bibinfo{author}{\bibfnamefont{F.}~\bibnamefont{Rom\'a}} \bibnamefont{et~al.},
  \bibinfo{journal}{Physica A} \textbf{\bibinfo{volume}{363}},
  \bibinfo{pages}{327} (\bibinfo{year}{2006}{\natexlab{b}}).

\bibitem[{\citenamefont{Rom\'a et~al.}(2009{\natexlab{a}})\citenamefont{Rom\'a,
  Risau-Gusman, Ramirez-Pastor, Nieto, and Vogel}}]{Roma2009a}
\bibinfo{author}{\bibfnamefont{F.}~\bibnamefont{Rom\'a}},
  \bibinfo{author}{\bibfnamefont{S.}~\bibnamefont{Risau-Gusman}},
  \bibinfo{author}{\bibfnamefont{A.~J.} \bibnamefont{Ramirez-Pastor}},
  \bibinfo{author}{\bibfnamefont{F.}~\bibnamefont{Nieto}}, \bibnamefont{and}
  \bibinfo{author}{\bibfnamefont{E.~E.} \bibnamefont{Vogel}}
  (\bibinfo{year}{2009}{\natexlab{a}}), \bibinfo{note}{in preparation}.

\bibitem[{\citenamefont{Rom\'a et~al.}(2009{\natexlab{b}})\citenamefont{Rom\'a,
  Bustingorry, and Gleiser}}]{Roma2009c}
\bibinfo{author}{\bibfnamefont{F.}~\bibnamefont{Rom\'a}},
  \bibinfo{author}{\bibfnamefont{S.}~\bibnamefont{Bustingorry}},
  \bibnamefont{and} \bibinfo{author}{\bibfnamefont{P.~M.}
  \bibnamefont{Gleiser}} (\bibinfo{year}{2009}{\natexlab{b}}),
  \bibinfo{note}{in preparation}.

\bibitem[{\citenamefont{{Le Ca\"er}}(1989{\natexlab{a}})}]{lecaer1989}
\bibinfo{author}{\bibfnamefont{G.}~\bibnamefont{{Le Ca\"er}}},
  \bibinfo{journal}{J. Phys. A} \textbf{\bibinfo{volume}{22}},
  \bibinfo{pages}{L647} (\bibinfo{year}{1989}{\natexlab{a}}).

\bibitem[{\citenamefont{{Le Ca\"er}}(1989{\natexlab{b}})}]{lecaer1989b}
\bibinfo{author}{\bibfnamefont{G.}~\bibnamefont{{Le Ca\"er}}},
  \bibinfo{journal}{Physica A} \textbf{\bibinfo{volume}{159}},
  \bibinfo{pages}{329} (\bibinfo{year}{1989}{\natexlab{b}}).

\bibitem[{\citenamefont{Binder and Luijten}(2001)}]{binder2001}
\bibinfo{author}{\bibfnamefont{K.}~\bibnamefont{Binder}} \bibnamefont{and}
  \bibinfo{author}{\bibfnamefont{E.}~\bibnamefont{Luijten}},
  \bibinfo{journal}{Phys. Rep.} \textbf{\bibinfo{volume}{344}},
  \bibinfo{pages}{179} (\bibinfo{year}{2001}).

\bibitem[{\citenamefont{Hukushima and Nemoto}(1996)}]{Hukushima1996}
\bibinfo{author}{\bibfnamefont{K.}~\bibnamefont{Hukushima}} \bibnamefont{and}
  \bibinfo{author}{\bibfnamefont{K.}~\bibnamefont{Nemoto}},
  \bibinfo{journal}{J. Phys. Soc. Jpn.} \textbf{\bibinfo{volume}{65}},
  \bibinfo{pages}{1604} (\bibinfo{year}{1996}).

\bibitem[{\citenamefont{Campbell and Bernardi}(1994)}]{campbell1994}
\bibinfo{author}{\bibfnamefont{I.~A.} \bibnamefont{Campbell}} \bibnamefont{and}
  \bibinfo{author}{\bibfnamefont{L.}~\bibnamefont{Bernardi}},
  \bibinfo{journal}{Phys. Rev. B} \textbf{\bibinfo{volume}{50}},
  \bibinfo{pages}{12643} (\bibinfo{year}{1994}).

\bibitem[{\citenamefont{Jan and Ray}(1994)}]{jan1994}
\bibinfo{author}{\bibfnamefont{N.}~\bibnamefont{Jan}} \bibnamefont{and}
  \bibinfo{author}{\bibfnamefont{T.~S.} \bibnamefont{Ray}},
  \bibinfo{journal}{J. Stat. Phys.} \textbf{\bibinfo{volume}{75}},
  \bibinfo{pages}{1197} (\bibinfo{year}{1994}).

\end{thebibliography}

\end{document}